\documentclass[twocolumn,chaos,aps,showpacs,showkeys]{revtex4-1}
\usepackage{graphicx,latexsym,amsfonts,amsmath,amssymb,bm}

\usepackage{amssymb,amsmath,amsthm}
\usepackage{graphics}
\usepackage{verbatim}
\usepackage{color}%

\newcommand{\be}{\begin{equation}}
\newcommand{\ee}{\end{equation}}
\newcommand{\ben}{\begin{eqnarray}}
\newcommand{\een}{\end{eqnarray}}

\begin{document}

\title{Inferring long memory processes in the climate network via
ordinal pattern analysis}

\author{Marcelo \surname{Barreiro}}
\author{Arturo C. \surname{Marti}}
\affiliation{Instituto  de F\'\i sica, Facultad de Ciencias,
Universidad de la Rep\'ublica, Igu\'a 4225, Montevideo, Uruguay}

\author{Cristina \surname{Masoller}}
\affiliation{Departament de Fisica i Enginyeria Nuclear, Universitat Politecnica de Catalunya, Colom 11,
E-08222 Terrassa, Barcelona, Spain}

\date{\today}

\begin{abstract}
We use ordinal patterns and symbolic analysis to construct global
climate networks and uncover long and short term memory processes. The
data analyzed is the monthly averaged surface air temperature (SAT
field) and the results suggest that the time variability of the SAT
field is determined by patterns of oscillatory behavior that repeat from
time to time, with a periodicity related to intraseasonal oscillations
and to El Ni\~{n}o on seasonal-to-interannual time scales.


\end{abstract}

\maketitle

{\bf
We analyze climatological data from a complex networks perspective, using techniques of nonlinear time-series symbolic analysis. Specifically, we employ ordinal patterns and binary representations to analyze monthly-averaged surface air temperature (SAT) anomalies. By computing the mutual information of the time-series in regular grid points covering the Earth's surface and then performing global thresholding, we construct climate networks which uncover short-term memory processes, as well as long ones (5-6 years). Our results suggest that the time variability of the SAT anomalies is determined by patterns of oscillatory behavior that repeat from time to time, with a periodicity related to intraseasonal variations and to El Ni\~no on seasonal to interannual time scales.
The present  work is located at the triple intersection of three highly
active interdisciplinary research fields in nonlinear science: symbolic methods
for nonlinear time series analysis, network theory, and nonlinear processes in the earth climate. While
a lot of effort is being done in order to improve our understanding of natural complex systems, with many
different methods for mapping time series to network representations
being investigated and employed in complex systems such as the human
brain, our work is the first one aimed at characterizing the global
climate network in terms of oscillatory patterns that tend to repeat
from time to time, with various time scales. By mapping these processes
into a global network, using ordinal patterns and binary
representations, we find that the structure of the network changes
drastically at different time scales.}

\maketitle

\section{Introduction}

Complex networks have been intensively studied in the last years because
they represent many real systems such as the Internet, ecological,
social and metabolic networks, genes, cells and the brain
\cite{sincro_comp_net}. Global climate modeling is also a hot topic
nowadays because of its huge economic and social impact for future
generations. Giving the complexity of the inter-relations between the
different elements that constitute our environment, it is important to
analyze climatological data from a complex network perspective. However,
despite the intensive effort in research done in these two
interdisciplinary and fascinating fields, just very few studies have
combined both
\cite{tsonis_bams,yamasaki_prl,tsonis_prl,tsonis_climate,kurths_epjst,kurths_epl,lehnertz_chaos}.
These studies have shown that network theory can yield light into interesting,
previously unknown features of our climate.

Tsonis and Swanson \cite{tsonis_prl} and Yamasaki, Gozolchiani and Halvin \cite{yamasaki_prl} have shown that the climate network is significantly affected by El Ni\~no, as during El Ni\~no years many links of the network are broken. Tsonis and Swanson \cite{tsonis_prl} constructed cross-correlation-based networks of the SAT field for El Ni\~no and for La Ni\~na years and investigated their structure. They found that the El Ni\~no network possesses significantly fewer links and lower clustering coefficient and characteristic path length than the La Ni\~na network. They conjectured and verified that, because El Ni\~no network is less communicative and less stable than La Ni\~na one, during El Ni\~no years temperature predictability is lower compared to La Ni\~na years. Using a different approach, Yamasaki, Gozolchiani and Halvin \cite{yamasaki_prl} arrived at a similar conclusion. They developed a method which allows to follow time variations of the network structure by observations of fluctuations in the correlations between nodes. The method allows to distinguish between the two qualitatively different groups of network links, blinking links that appear and disappear in a short time, and robust links that represent long lasting relations between temperature fluctuations in two regions. Assuming that broken links are due to structural changes in the network, by tracking these changes in several zones a strong response to El Ni\~no was reveled, even in geographical regions where the mean temperature is not affected by El Ni\~no.

Donges {\it et al.} \cite{kurths_epjst} compared the structural properties of networks constructed by using, as a measure of dynamical similarity between regions, linear and nonlinear measures: the linear Pearson correlation coefficient and the nonlinear mutual information. They analyzed two sets of data: the SAT anomalies obtained from large-scale climate simulations by the coupled atmosphere-ocean general circulation models and the SAT anomalies reanalysis data sets. A high degree of similarity using the two approaches (linear and nonlinear similarity measures) was found on the local and on the mesoscopic topological scales; however, important differences were uncovered on the global scale, particularly in the betweenness centrality field. In \cite{kurths_epl} Donges {\it et al.} employed the mutual information to reveal wave-like structures of high-energy flow, that could be traced back to global surface ocean currents. Their results point to the major role of the oceanic surface circulation in coupling and stabilizing the global temperature field in the long-term mean.

When computing the mutual information, in order to detect patterns and correlations in the variability of two
nodes, a critical issue is defining probability distribution functions (PDFs) that fully take into
account the temporal order in which the SAT anomalies occur in the time-series. Histogram-based PDFs do not take into account this temporal order, and thus, are not optimal for capturing subtle correlated oscillatory patterns. Alternatively, one can
use time-delay embedding techniques to represent the time series as a trajectory in
a high-dimensional space; however, the information provided by the
mutual information is strongly dependent on the embedding technique, the
time-delay, and the phase space partition \cite{grassberger}.

An alternative methodology, originally proposed by Bandt and
Pompe (BP) \cite{Pompe02} allows to define probability distribution functions that fully take
into account the time ordering of the SAT anomalies. The BP method is based on comparing values in the time-series to construct ''ordinal patterns''. By computing the PDF of the possible ordinal patterns, various information-theory quantifiers, such as the permutation entropy, the mutual information, complexity measures, etc. can be computed. The BP method has been successfully employed to analyze time-series generated from physical, biological and social systems (see, e.g., \cite{amigo_book} and references therein).

When employing the BP methodology the precise values of the SAT anomalies are neglected (as the method is based on comparing relative values in the time-series); however, as we will show, with the BP method one can identify patterns of oscillatory behavior that tend to repeat from time to time, with various time scales. A drawback of the BP method is that, in order to capture long memory processes, long time series are needed to compute the PDFs of the ordinal patterns with good statistics. The SAT data available (described in the next section) limited us to construct ordinal patterns of maximum length 5, which allows to consider time-scales up to 5 years or 5 months. To overcome this limitation we employed ``binary representations", by which the time-series of SAT anomalies were transformed into sequences of 0s and 1s. These binary representations allowed to consider processes with longer time-scales, up to 6 years or 6 months. We will show in what follows that ordinal patterns and binary representations are tools that, when employed within a complex network perspective, are very powerful for the analysis of climatological data. By reveling long term and short term memory processes they provide additional information to that obtained from conventional time-series analysis and thus they help to a better understanding of our complex climate.

This article is organized as follows: Section II presents the description of the data analyzed and a summary of the methodology employed. Section III presents the results obtained with ordinal patterns and binary representations, and a comparison with the methodologies previously employed by other authors (i.e., the linear cross-correlation \cite{tsonis_prl} and the nonlinear histogram-based mutual information \cite{kurths_epjst}).  Section IV contains a discussion of the results and the conclusions.

\section{Data and methods}

We present the analysis of the monthly averaged surface air temperature
(SAT field, reanalysis data from the National Center for Environmental
Prediction/National Center for Atmospheric Research, NCEP/NCAR \cite{kalnay}). As in
\cite{tsonis_bams,tsonis_prl,kurths_epjst,kurths_epl}, anomaly values are
considered (i.e., the actual temperature value minus the monthly average).

The data covers a regular grid over the earth's surface with latitudinal
and longitudinal resolution of 2.5$^0$. These $N=10226$ grid points are
considered the nodes (or vertices) of a network (or graph), and the
existence of a link (or edge) between any two nodes depends on the
``weight" of the link that measures the degree of statistical similarity
between the climate dynamics in those two nodes. The data covers the
period January 1949-December 2006, and therefore in each grid point $i$
($i=1\dots N$) we have $M=696$ data points, $\{x_i(t), t=1\dots
M\}$. $W=\{w_{ij}, i,j=1\dots N\}$ is the matrix that contains the
weights that characterize the links between any two nodes. Since we
don't attempt to uncover directionality in the couplings among the
nodes, we will consider a symmetric measure of statistical similarity
that results in symmetric weights.

In \cite{tsonis_bams,tsonis_prl} these weights were quantified with the
absolute value of the linear cross-correlation coefficient; in
\cite{kurths_epjst,kurths_epl}, with the mutual information, a nonlinear
measure that is a function of the probability density functions (PDFs)
that characterize the time series in the two nodes, $p_i(m)$ and
$p_j(n)$, as well as of the joint probability, $p_{ij}(m,n)$,
\begin{equation}
\label{mutual}
W_{ij}=M_{ij}=\sum_{m,n} p_{ij}(m,n) \log {{p_{ij}(m,n)}\over{p_{i}(m)p_{j}(n)}}.
\end{equation}
The mutual information, which can also be written as
\begin{equation}
M_{ij}=S_i + S_j - S_{ij},
\end{equation}
where $S_i=-\sum p_i \log p_i$, $S_j=-\sum p_j \log p_j$ and $S_{ij}=-\sum p_{ij} \log p_{ij}$, indicates the amount of information of  $\{x_i(t)\}$, we obtain by knowing  $\{x_j(t)\}$, and vice versa. $M_{ij}$ measures the degree of statistical interdependence of the time series; if they are independent, $p_{ij}(m,n)=p_i(m)p_j(n)$ and $M_{ij}=0$.

To uncover correlated ``patterns" of oscillatory behavior in the SAT anomalies, we employ the
methodologies referred to as ordinal patterns and symbolic analysis,
which are based on comparing consecutive values in the time series, to compute the PDFs in Eq.(\ref{mutual}). We
begin by presenting the ordinal pattern methodology \cite{Pompe02}.

First, in each grid point $i$, the time series $\{x_i(t) \}$ is divided
into $M-D$ overlapping vectors of dimension $D$. Then, each element of a
vector is replaced by a number from 0 to $D-1$, in accordance with its
relative magnitude in the ordered sequence (0 corresponding to the
smallest and $D-1$ to the largest value in each vector). For example,
with $D=3$ the vector $(v_0,v_1,v_2)=(6.8, ~11.5, ~1.1)$, gives the
ordinal pattern $201$ because $v_2 < v_0 < v_1$. In this way, each
vector has associated an ``ordinal pattern" (OP) composed by $D$
symbols, and the symbol sequence comes from a comparison of neighboring
values. Last, one computes the PDF of the $D!$ possible ordinal
patterns. For example, with $D=3$ the $3!=6$ different patterns are
(012, 021, 102, 120, 201 and 210), and thus, the PDF is calculated with
6 bins. To have a good statistics one must have $M-D>>D!$ (i.e., \# of
OPs in the time series $>>$ \# of possible OPs).

Because in each time series we have $M=696$ data points,
to compute the PDFs with good statistics we limit to consider only $D=4$ and $D=5$.
Ordinal patterns of $D\le3$ do not provide good resolution for computing the mutual
information, Eq. \ref{mutual}, because the PDFs are calculated with very few bins (for $D=6$, there are only 6 ordinal patterns, and thus, only 6 bins).

With climatological data meaningful ordinal patterns can be formed
either by comparing consecutive years or consecutive
months. Specifically, if we use $D=3$, when comparing consecutive years,
the OPs in node $i$ are defined by ($x_i(t),x_i(t+12), x_i(t+24))$, $t=1,\dots, M-24$; when comparing consecutive
months, they are defined by ($x_i(t), x_i(t+1), x_i(t+2))$, $t=1,\dots,
M-2$.

To decide whether there is a link between two nodes, we perform global
thresholding \cite{victor}, i.e., we define a threshold $\tau$ (which
is the same for all pairs of nodes) and assume that there is a link
between $i$ and $j$ if the weight of the link is above the threshold,
i.e., $w_{ij}\ge \tau$.

Clearly, a careful selection of the threshold is crucial for uncovering the backbone of the network \cite{kurths_epjst}.

We use the following procedure: first, we check that we only take into account significant network connections. To do this we compute the weight matrix $W$ from randomly shuffled time series in each node. The random elements of this 10226$\times$10226 matrix have a very narrow PDF which, in principle, allows the use of the maximum matrix element, $w_{max}$, as a significant limit. Then, we compute $W$ with the original time series and consider that there is a significant link between the nodes $i$ and $j$ if $w_{ij}>w_{max}$, otherwise, we set $w_{ij}=0$. While there are several methods to eliminate non-significant links and the evaluation of statistical significance is still an open problem (see, e.g., the discussion in \cite{kurths_epjst}), this procedure is computationally cost-efficient and we will show in what follows that allows to uncover meaningful climate networks. The drawback is that it is a rather strong test that eliminates weak but significant links, and as a results the networks tend to be very spare. The final step is to chose a threshold $\tau$ to select the strongest links. As in \cite{kurths_epjst}, we chose $\tau$ such that the resulting networks have a pre-determined number of links. In the following we present results for networks that have 1\% of the total possible links (which will be referred to as ``low-threshold'' networks) and networks containing 0.1\% of the total possible links (referred to as ``high-threshold'' networks). For easy comparison and to visualize the effect of thresholding, we also present the networks containing {\it all the significant links}, which will be referred to as the ``zero-threshold'' networks.

The networks are represented graphically as two-dimensional maps by
plotting the {\it area-weighted connectivity}
\cite{tsonis_bams,tsonis_prl,kurths_epjst,kurths_epl}, which is the
fraction of the total area of the earth to which each node $i$ is
connected,
 \begin{equation}
\label{awc}
AWC_{i}={{\sum_{j}^{N} A_{ij}\cos(\lambda_i)}\over{\sum_{j}^{N} \cos(\lambda_j)}},
\end{equation}
where $\lambda_i$ is the latitude of node $i$ and $A_{ij}=1$ if nodes
$i$ and $j$ are connected (i.e., if $w_{ij}\ge \tau$) and 0 otherwise.
The cosine terms correct for the fact that in a surface spherical
network defined on a regular planar grid, the nodes correspond to
regions of different area.

\graphicspath{{c:/fortran/clima/temp/}}
\begin{figure*}[htbn]
\begin{center}
\includegraphics[width=2.0\columnwidth]{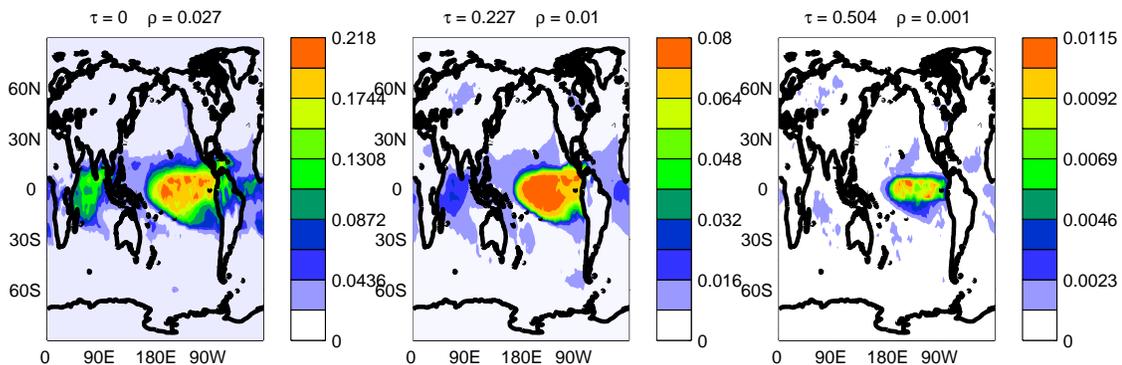}
\end{center}
\caption{Zero-threshold (left), low-threshold (center) and high-threshold (right) networks
constructed by computing the mutual information from ordinal patterns of length $D=4$ defined by comparing SAT anomalies in
consecutive years. The 2D plots are color-coded such that the white (red) regions indicate the geographical areas with {\it zero} (largest) area weighted connectivity. In each panel the values of the threshold, $\tau$, and of the edge-density, $\rho$, are indicated.}
\label{fig:bp4a}
\end{figure*}
\begin{figure*}[htbn]
\begin{center}
\includegraphics[width=1.4\columnwidth]{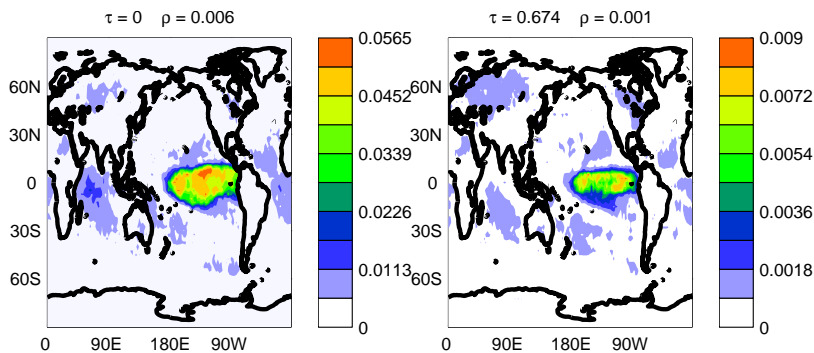}
\end{center}
\caption{Zero-threshold (left) and high-threshold (right) networks
constructed by computing the mutual information from ordinal patterns of length $D=5$ defined by comparing SAT anomalies in
consecutive years. The 2D plots are color-coded such that the white (red) regions indicate the geographical areas with {\it zero} (largest) area weighted connectivity. The weaker links lose a bit of memory (compare the zero-threshold networks with $D=4$ and $D=5$) while the strong links do not, as the high-threshold networks are the nearly same for $D=4$ and $D=5$.}
\label{fig:bp5a}
\end{figure*}

\section{Results}

\subsection{Ordinal Patterns Analysis}
\begin{figure*}[htbn]
\begin{center}
\includegraphics[width=2.0\columnwidth]{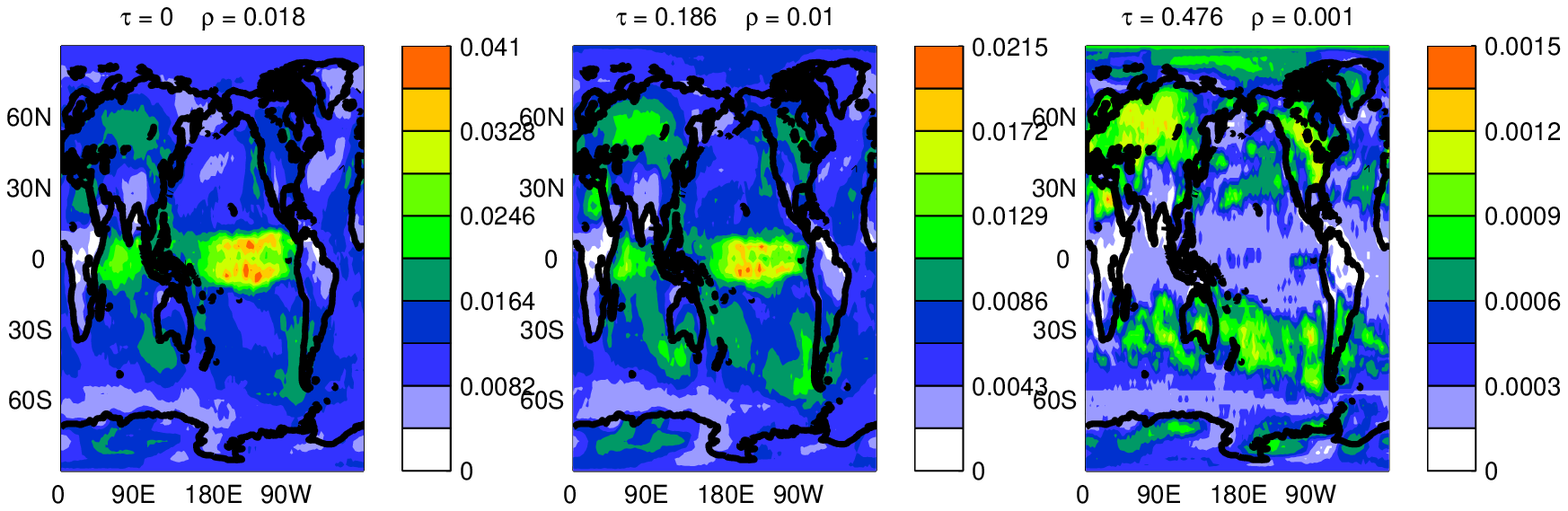}
\end{center}
\caption{As Fig. \ref{fig:bp4a} but $D=4$ ordinal patterns defined by comparing SAT anomalies in consecutive months. The 2D plots of the area weighted connectivity are color-coded such that the white (red) regions indicate the geographical areas with {\it zero} (largest) area weighted connectivity.}
\label{fig:bp4m}
\end{figure*}

\begin{figure*}[htbn]
\begin{center}
\includegraphics[width=1.4\columnwidth]{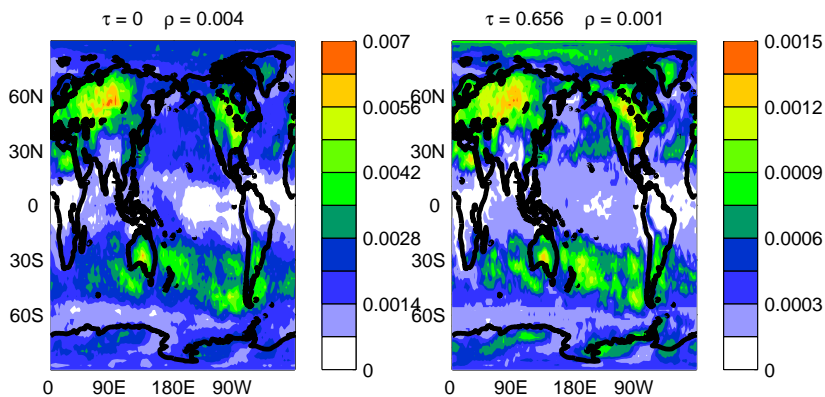}
\end{center}
\caption{Zero-threshold (left) and high-threshold (right) networks
constructed by computing the mutual information from ordinal patterns of length $D=5$ defined by comparing SAT anomalies in
consecutive months. The 2D plots of the area weighted connectivity are color-coded such that the white (red) regions indicate the geographical areas with {\it zero} (largest) area weighted connectivity.}
\label{fig:bp5m}
\end{figure*}

The networks obtained when the ordinal patterns are defined by comparing
SAT anomalies in consecutive years and in consecutive months are displayed
in Figs. \ref{fig:bp4a}-\ref{fig:bp5m}. In each panel the values of the threshold, $\tau$, and of the edge-density,
\begin{equation}
\rho = {{\sum_{i,j}^N A_{ij}}\over{N(N-1)}},
\end{equation}
are indicated.

For consecutive years the networks with $D=4$, Fig. \ref{fig:bp4a}, and $D=5$, Fig. \ref{fig:bp5a} are
very similar showing highest connectivity for the tropical region.


For zero-threshold (left panels in Figs. \ref{fig:bp4a} and \ref{fig:bp5a}) the tropical Pacific shows the largest
connectivity, particularly in the central and eastern side of the basin;
the tropical Atlantic and Indian oceans follow. In the extratropics
there are patches of high connectivity off the western coast of Canada
in the Northern Hemisphere (N.H.)  and in the south Pacific in the
Southern Hemisphere (S.H.).  This connectivity structure is more pronounced for $D=5$, although some of the weak links
lose memory.  These characteristics hint to El Ni\~{n}o  as a
fundamental player in setting up these connections \cite{clima1}.  The
El Ni\~{n}o phenomenon occurs on interannual time scales and consists in
an anomalous warming of the eastern equatorial Pacific. This warming in
turn warms up the local atmosphere and influences other tropical regions
through the excitation of equatorial Kelvin and Rossby waves. Changes in
the precipitation associated with El Ni\~{n}o also induce stationary
Rossby waves in the northern and southern extratropics that generate
long range connections called atmospheric teleconnection patterns.
Examples of these structures are the Pacific-North American pattern that
affects the northern Pacific and North America, and the Pacific-South
American pattern that propagates in the southern Pacific toward South
America. These anomalous structures connect the tropical Pacific with
remote locations and affect the local climate by changing, for example,
the advection of heat or moisture into a region.

For non-zero-thresholds (center and right panels in Figs. \ref{fig:bp4a}
and \ref{fig:bp5a}) only the strongest links remain and the networks clearly
show again an El Ni\~{n}o-like structure in the tropical Pacific.
Secondary maxima in the Indian and Atlantic oceans are present in the low-threshold
networks but are significantly weakened in the high-threshold ones. The continental regions have overall very low connectivity which translates in the low predictability of surface temperature anomalies on interannual time scales. Within these continental regions, the largest connectivity is seen over Asia and North America, the latter maximum being perhaps due to the Pacific North American pattern induced by El Ni\~{n}o \cite{pna}.

The networks obtained when the ordinal patterns are defined by comparing
temperature anomalies in consecutive months, Figs. \ref{fig:bp4m}, \ref{fig:bp5m}, present, for zero- and low-
threshold, similar features as for consecutive years,
although the networks are more homogeneous. There is a maximum in the
equatorial Pacific, a secondary maximum in the Indian ocean and
extratropical maxima over Asia, North America and southern
subtropics. On the other hand, the high threshold network shows that the strongest links are located in the extra tropics. We speculate that this could be a result of the modulation of the temperature variance by the seasonal cycle, which is strongest over the northern hemisphere continental masses and has a minimum in the tropical band. This network structure is also seen when using ordinal patterns formed by 5 consecutive months, Fig. \ref{fig:bp5m}.

\subsection{Binary representations}

To capture longer memory processes one should use larger $D$ values;
however, for $D = 6$ there are 6! = 720 possible ordinal patterns, and
since we have time series with less than 700 data points, there is not
enough data to calculate ordinal patterns PDFs with good statistics.

\begin{figure*}[htbn]
\begin{center}
\includegraphics[width=1.9\columnwidth]{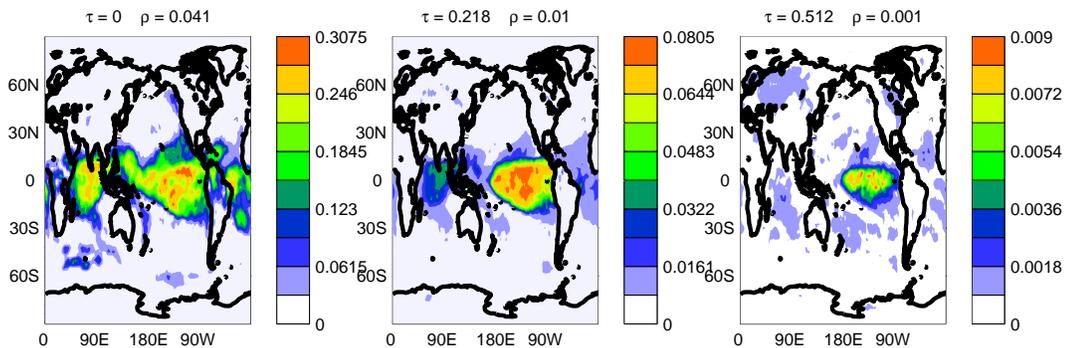}
\end{center}
\caption{As Fig. \ref{fig:bp4a} but employing binary representation. Zero-threshold (left), low-threshold (center) and high-threshold (right) networks constructed by computing the mutual information from patterns of length $D=4$ defined by comparing SAT anomalies in
consecutive years. The 2D plots are color-coded such that the white (red) regions indicate the geographical areas with {\it zero} (largest) area weighted connectivity. In each panel the values of the threshold, $\tau$, and of the edge-density, $\rho$, are indicated.}
\label{fig:bin4a}
\end{figure*}
\begin{figure*}[htbn]
\begin{center}
\includegraphics[width=1.9\columnwidth]{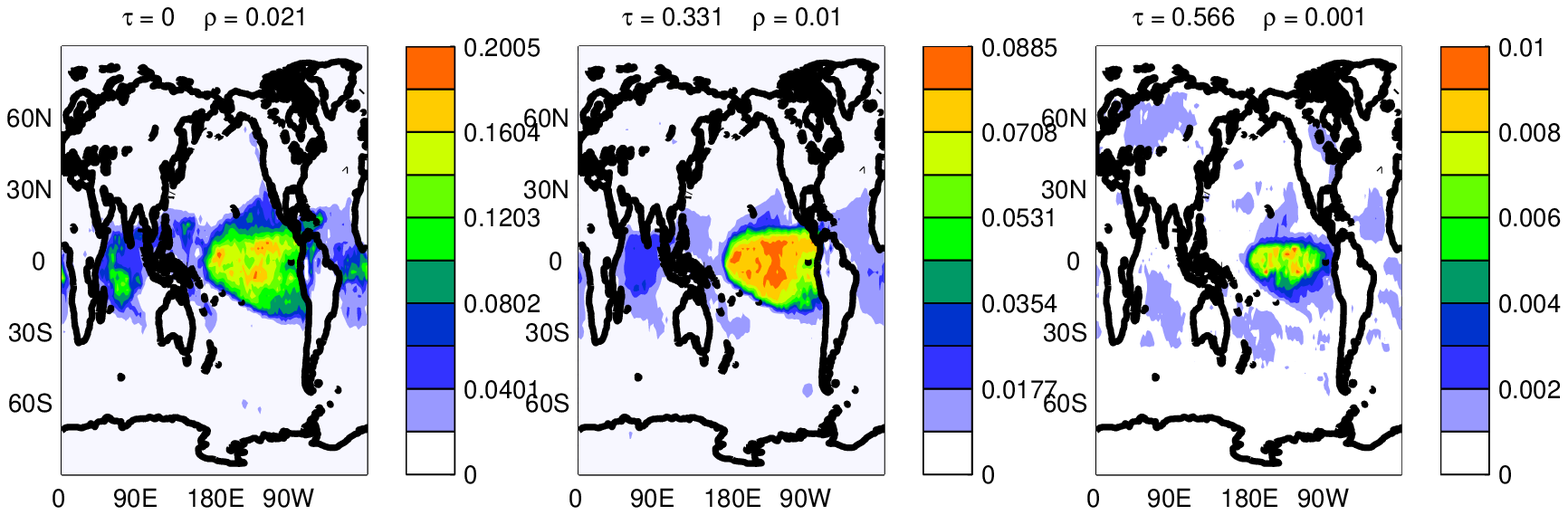}
\end{center}
\caption{As Fig. \ref{fig:bin4a} but with $D=5$. }
\label{fig:bin5a}
\end{figure*}
\begin{figure*}[htbn]
\begin{center}
\includegraphics[width=1.3\columnwidth]{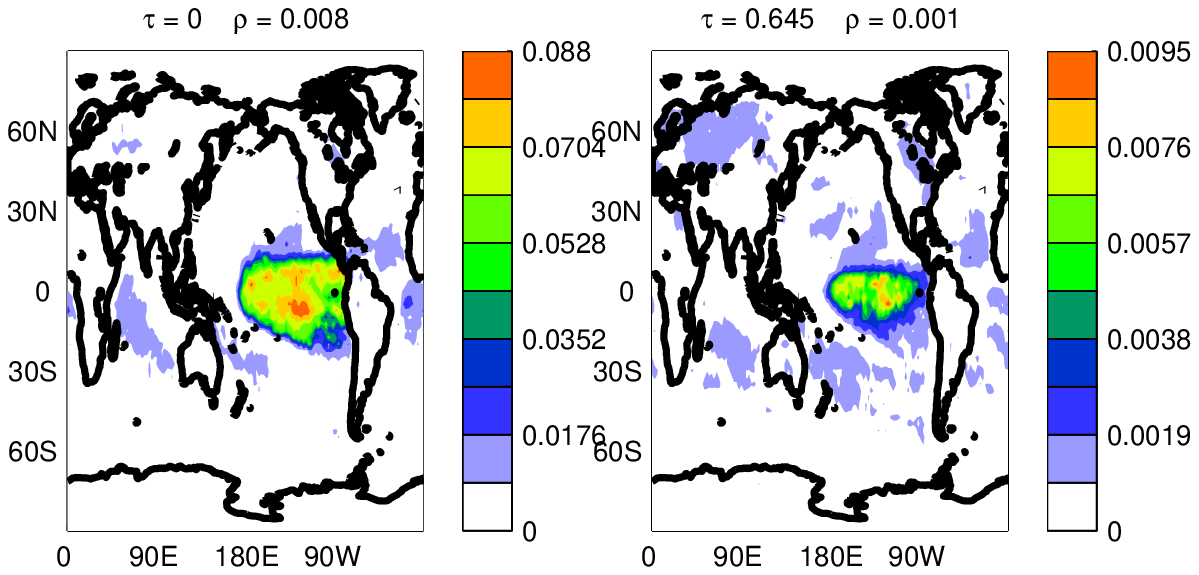}
\end{center}
\caption{As Fig. \ref{fig:bin4a} but with $D=6$. Comparing with Figs. \ref{fig:bin4a} and \ref{fig:bin5a} one can see that there is a good agreement with the results obtained previously with ordinal patterns: the weaker links tend to lose memory, while the strongest links do not.}
\label{fig:bin6a}
\end{figure*}

As discussed in the introduction, a solution to overcome this problem is employing ``binary representations", by
which the time series $\{x_i(t)\}$ is transformed into a sequence
$\{v_i(t)\}$ of 0s and 1s, using the following rule: $v_i(t)=0$ if
$x_i(t)\ge 0$ and $v_i(t)=1$ otherwise (since the $x_i$ values are
temperature anomalies, we are taking into account whether the SAT field
is above or below its monthly averaged value). We can then define
``binary patterns" of dimension $D$ (e.g., for $D=3$ the possible
patters are 000, 001, 010, 100, 011, 110, 101 and 111) and compute their
PDF. The number of different patterns is $2^D$, and thus, we can
calculate PDFs of patterns of $D=6$ ($2^6=64$) with good enough statistics.
Patterns with $D\le3$ do not provide good resolution for computing the mutual
information, Eq. \ref{mutual}, because the PDFs are calculated with very few bins.
Therefore, in the following, we consider $D=4$, 5, and 6.

\begin{figure*}[htbn]
\begin{center}
\includegraphics[width=2.0\columnwidth]{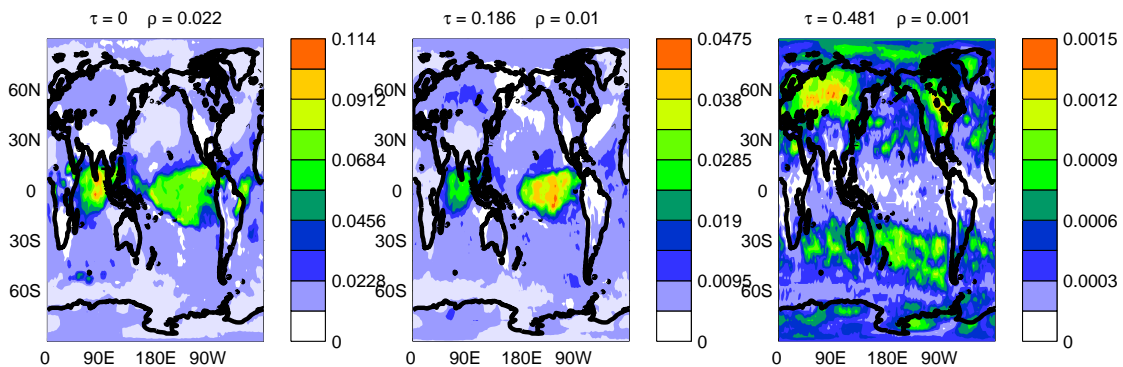}
\end{center}
\caption{As Fig.\ref{fig:bin4a} but the networks constructed with binary representation comparing anomalies in $D=4$ consecutive months. The 2D plots of the area weighted connectivity are color-coded such that the white (red) regions indicate the geographical areas with {\it zero} (largest) area weighted connectivity.}
\label{fig:bin4m}
\end{figure*}

\begin{figure*}[htbn]
\begin{center}
\includegraphics[width=1.4\columnwidth]{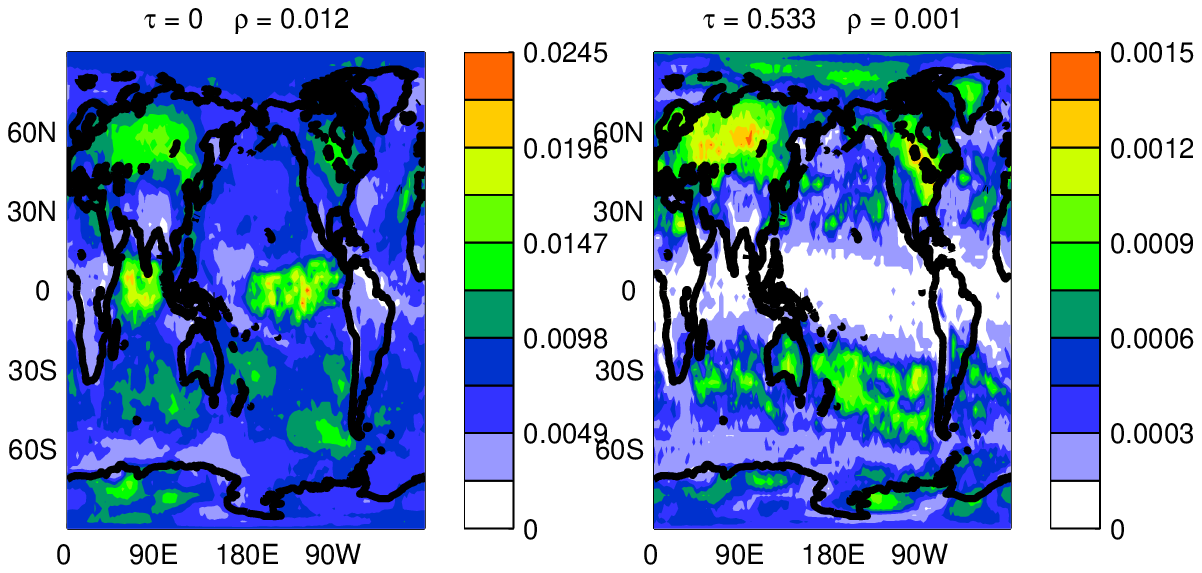}
\end{center}
\caption{As Fig. \ref{fig:bin4m} but with $D=5$. }
\label{fig:bin5m}
\end{figure*}

\begin{figure*}[htbn]
\begin{center}
\includegraphics[width=1.4\columnwidth]{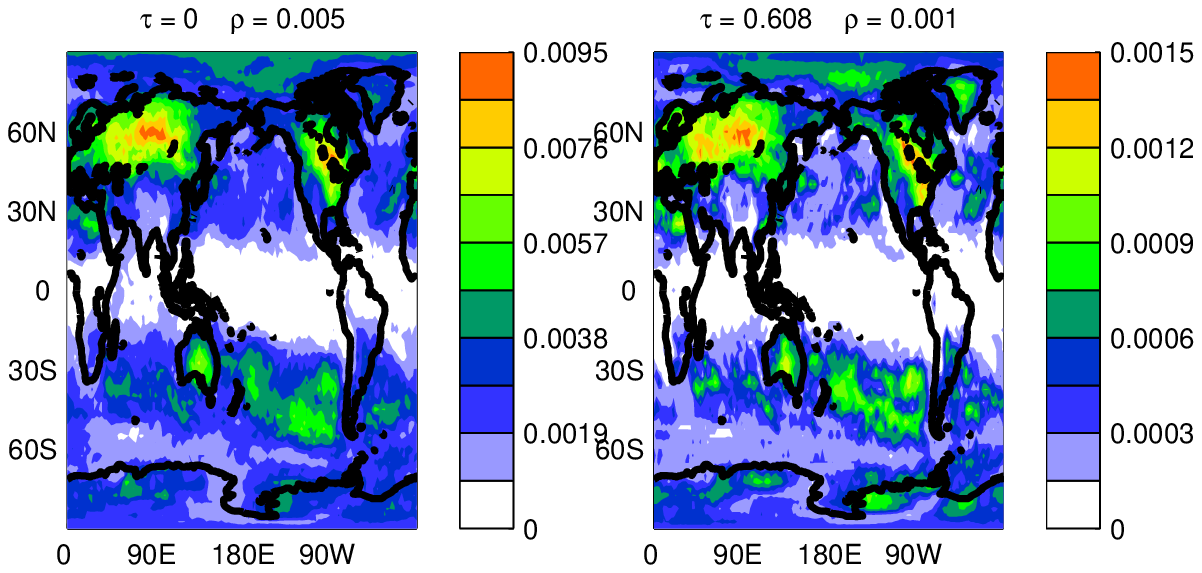}
\end{center}
\caption{As Fig. \ref{fig:bin4m} but with $D=6$.}
\label{fig:bin6m}
\end{figure*}

Figures \ref{fig:bin4a}-\ref{fig:bin6m} present the results when
the binary patterns are defined by consecutive years and months.


For consecutive years, Figs. \ref{fig:bin4a}-\ref{fig:bin6a}, the networks obtained when using binary representations
are very similar to those found with ordinal patterns. The tropical regions are quite uniformly well connected (although a Pacific maximum is clear) while the
extratropics show localized regions of high connectivity likely due to atmospheric teleconnections forced from the tropics, particularly for low density networks.

The networks obtained for consecutive months, Figs. \ref{fig:bin4m}-\ref{fig:bin6m}, show that as the threshold or as $D$
increases there are overall similar changes in structure as those seen for ordinal patterns, Figs. \ref{fig:bp4m}, \ref{fig:bp5m}: for short memory processes (or for low threshold) the maximum connectivity is in the tropics, while for
longer time scales (or for higher threshold) the extratropics show largest number of links. As discussed before, this could be the result of the modulation of the temperature variance by the seasonal cycle, which would be the main process that connects grid points in very low density networks (grid points that are connected by very strong links). Our results could also hint at the role of land surface conditions like snow or soil humidity in increasing the persistence of surface
temperature anomalies over the northern hemisphere continents. Overall, these results agree well with the fact
that temperature teleconnections from the tropical Pacific tend to last
no much longer than a season in the different parts of the world.

\begin{figure*}[tbh]
\begin{center}
\includegraphics[width=2.0\columnwidth]{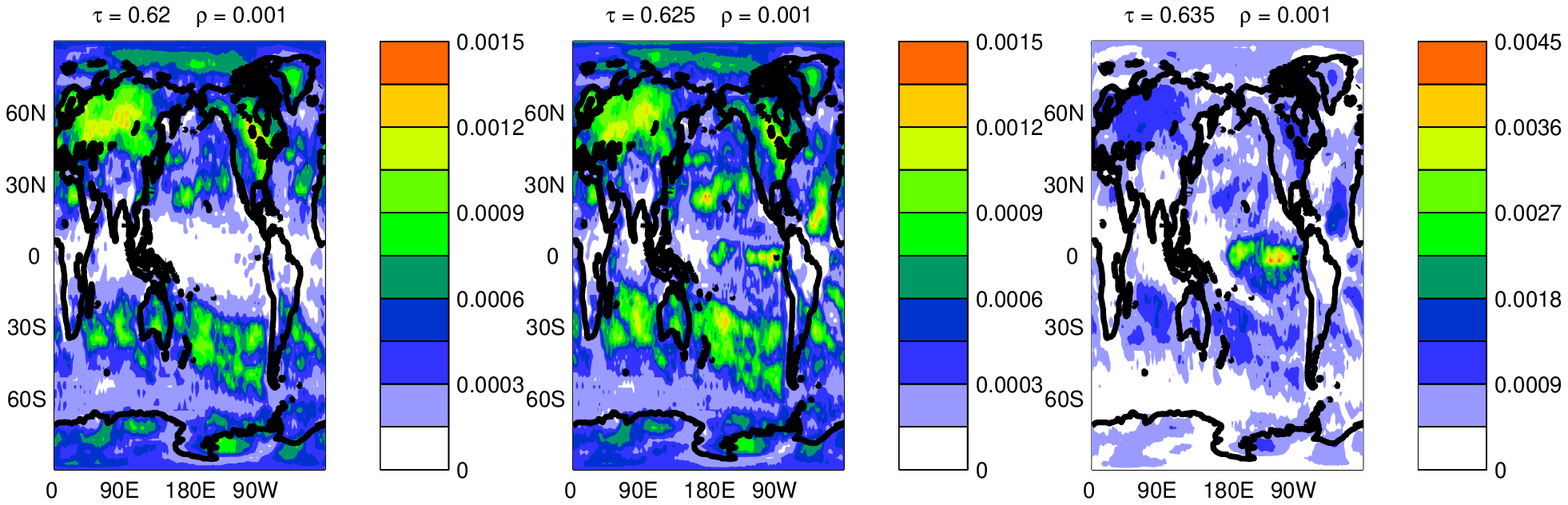}
\end{center}
\caption{High-threshold networks constructed with binary representations, with patterns of $D=6$ covering time-intervals of one year (left), two years (center) and three years (right). The 2D plots of the area weighted connectivity are color-coded such that the white (red) regions indicate the geographical areas with {\it zero} (largest) area weighted connectivity. See text for details.}
\label{fig:mix}
\end{figure*}
As a way to test the interpretation of the above presented results, in terms of the symbolic methodology of time-series analysis capturing two different time-scales of the Earth's climate, seasonal and interannual, we constructed binary patterns of fixed dimension, $D=6$, that cover three different time intervals:

i) covering one-year, the patterns are composed as
\begin{equation}
[x_i(t), x_i(t+2), x_i(t+4), \dots, x_i(t+10)];\nonumber
\end{equation}
ii) covering two-years, the patterns are composed as
\begin{equation}
[x_i(t), x_i(t+4), x_i(t+8), \dots, x_i(t+20)];\nonumber
\end{equation}
ii) covering three-years, the patterns are composed as
\begin{equation}
[x_i(t), x_i(t+6), x_i(t+12), \dots, x_i(t+30)].\nonumber
\end{equation}
The results are presented in Fig.{\ref{fig:mix}}, were one can see how the network changes. For a time interval of one year the extra tropics have the largest number of links and there are very few in the tropical region. On the other hand, for a time interval of three years, the extra-tropics keep about the same number of connections while in tropical Pacific ``El Ni\~{n}o'' stands out (note the different color scales in the panels in Fig.{\ref{fig:mix}}). In summary, confirming the results previously found with ordinal patterns and binary representations composed by consecutive years and by consecutive months, on intra-seasonal time scales the extra tropical connections dominate, while on interannual scales, the ``El Ni\~{n}o'' is the key player in setting up teleconnections worldwide.

\begin{figure*}[tbh]
\begin{center}
\includegraphics[width=2.0\columnwidth]{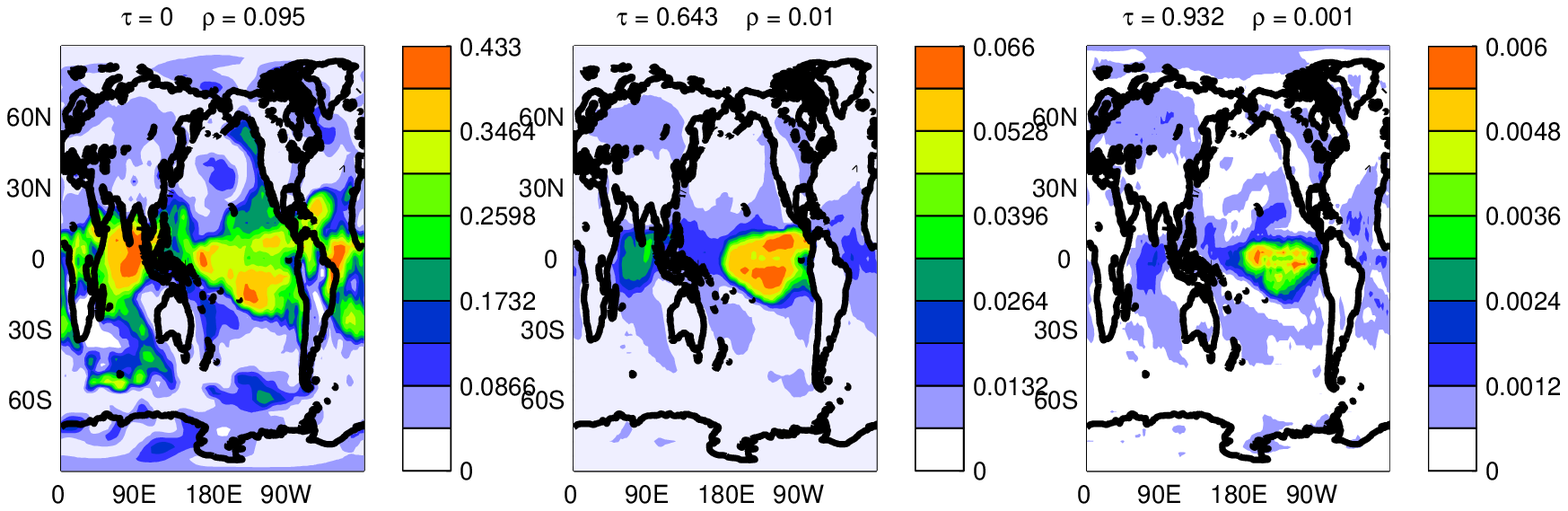}
\end{center}
\caption{Zero-threshold network (left) and non-zero-threshold networks (center and right)
constructed by estimating the weights with the absolute value of the cross-correlation coefficient. The 2D plots of the area weighted connectivity are color-coded such that the white (red) regions indicate the geographical areas with {\it zero} (largest) area weighted connectivity.}
\label{fig:cc}
\end{figure*}
\begin{figure*}[tbh]
\begin{center}
\includegraphics[width=2.0\columnwidth]{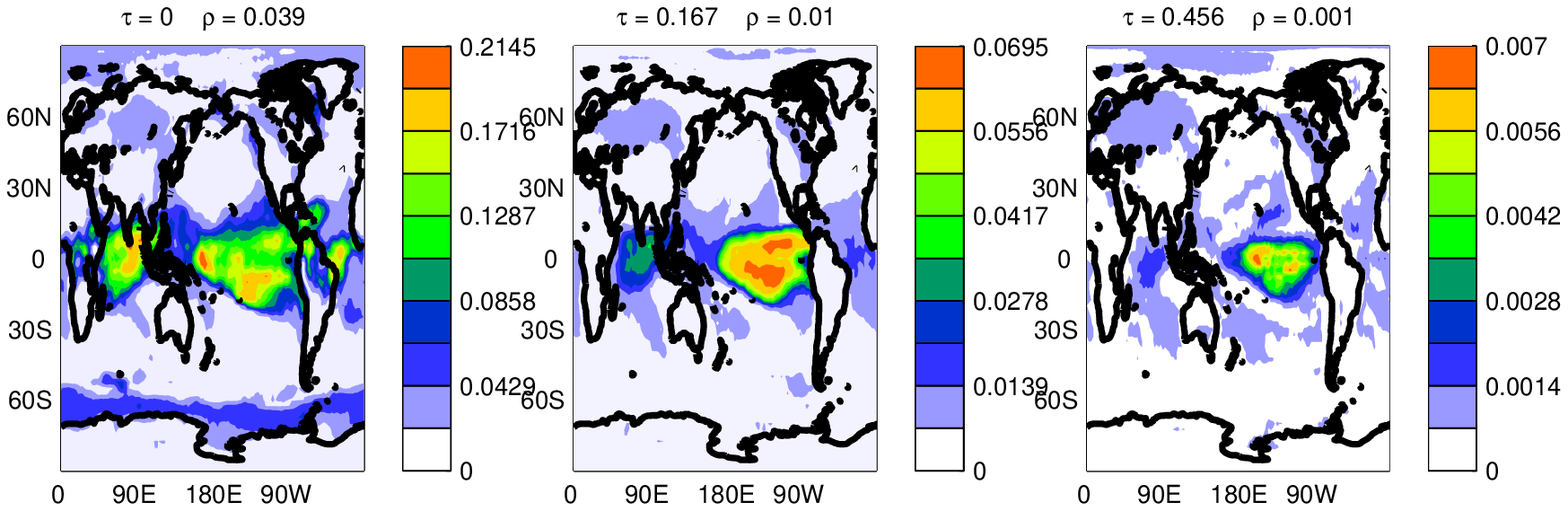}
\end{center}
\caption{Zero-threshold network (left) and non-zero-threshold networks (center and right)
constructed by estimating the weights with the mutual information, calculating the PDFs from histograms of SAT anomalies. The 2D plots of the area weighted connectivity are color-coded such that the white (red) regions indicate the geographical areas with {\it zero} (largest) area weighted connectivity.}
\label{fig:mm}
\end{figure*}

As it was previously discussed, the significance of the network links was tested in comparison with links computed from surrogate time-series in each node. Since surrogate data does not preserve the autocorrelation properties of the original time-series, to further test the validity of the previously presented results, we did the following test: we computed the mutual information using the original time series in node $i$ and the time-inverted series in node $j$. The results show no significant spatial structure in the area weighted connectivity plots (not shown).

\subsection{Comparison with other measures}

It is interesting to compare the results obtained using ordinal patterns and binary representations with those obtained using conventional techniques of time-series analysis, as the linear cross-correlation coefficient (as in Ref. \cite{tsonis_prl}) and the
mutual information, computing the PDFs from standard histograms of amplitude values (as in Refs. \cite{kurths_epjst,kurths_epl}). Figure \ref{fig:cc} displays the zero, low and high threshold networks when the weights are calculated with the absolute value of the cross-correlation coefficient coefficient and Fig. \ref{fig:mm}, when they are calculated with mutual information, with the PDFs calculated from histograms of temperature anomaly values. In this case the PDFs were computed employing 32 bins and in each time-series the values of the SAT anomalies were re-normalized such that each time-series has zero mean and standard deviation equal to one (as in Ref. \cite{kurths_epjst,kurths_epl}, for easier comparison).

The 2D plots of the area weighted connectivity are similar to those previously reported in Ref. \cite{kurths_epjst} and also, to those seen in Figs. \ref{fig:bp4a}, \ref{fig:bp5a}, \ref{fig:bin4a}, \ref{fig:bin5a}, where the ordinal patterns and the binary representations are formed by comparing consecutive years. ``El Ni\~{n}o'' is the main feature uncovered. There are also regions with relatively high number of links in the northern hemisphere continents and southern subtropics, but the high connectivity in the extratropics seen previously in Figs. \ref{fig:bp4m}, \ref{fig:bp5m}, \ref{fig:bin4m}, \ref{fig:bin5m}, and \ref{fig:bin6m} is not observed. In other words, employing the cross correlation coefficient or the histogram-based mutual information uncovers mainly the interannual network. These methodologies fail to separate the two distinct time-scales (intra-seasonal and inter-annual) that are clearly seen when using symbolic analysis and the time series are transformed in sequences of patterns by comparing consecutive years or consecutive months.

\section{Conclusions}

Concluding, we have shown that ordinal patterns and symbolic analysis
applied to anomalies of the surface air temperature are powerful tools
for the analysis of the large-scale topology of the climate network. The
success of these methods is based on an appropriate partition of the
phase space that results in ordinal patterns and binary representations
having PDFs that characterize the diversity of patterns present in the
climate dynamics.

A main advantage of the methodology proposed here is that by varying the
dimension of the pattern and the year-month comparison, one can uncover
memory processes with different time scales. We found that both, monthly
and yearly patterns reveal long memory processes, and that depending on
the time scale considered the climate network can change
completely.

The fact that ordinal patterns and symbolic analysis give meaningful
information indicates that the time variability of the anomaly SAT field
is strongly determined by patterns of oscillatory behavior that tend to
repeat from time to time.

Overall we found that on seasonal time-scales the extratropical regions,
mainly over Asia and North America, present the strongest links while in
interannual time scales, the tropical Pacific clearly dominates.

\begin{acknowledgments}
The authors thank the two anonymous referees for their very useful comments and suggestions. M.B. acknowledges support from the European Community's Seventh Framework Programme (FP7/2007-2013) under Grant Agreement N° 212492 (CLARIS LPB. A Europe-South America
Network for Climate Change Assessment and Impact Studies in La Plata Basin). C.M. acknowledges support from the ICREA Academia programme, the Ministerio de Ciencia e Innovaci\'on, Spain, project FIS2009-13360 and the Agencia de Gestio d'Ajuts Universitaris i de Recerca (AGAUR), Generalitat de Catalunya, through project 2009 SGR 1168.
\end{acknowledgments}

\end{document}